\documentclass[preprint]{ptephy_v1}
\usepackage{hyperref}

\newcommand\sped{$\sigma_{\rm ped}$\,\,}
\newcommand\Vbias{$V_{\rm bias}$\,\,}

\begin{document}

\title{Precision beam telescope based on SOI pixel sensor technology for electrons in the energy range of sub-GeV to GeV}


\author[1]{Hisanori Suzuki}
\author[1]{Takumi Omori}
\author[1]{Kazuhiko Hara}
\author[1]{Hiroki Yamauchi}
\affil{University of Tsukuba, Tsukuba, Ibaraki 305-8571, Japan}

\author[2]{Miho Yamada} 
\affil{Tokyo Metropolitan College of Industrial Technology, Arakawa, Tokyo 116-8523, Japan}

\author[3]{Toru Tsuboyama}
\affil{High Energy Accelerator Org. (KEK), Tsukuba, Ibaraki 305-0801}

\author[4]{Ayaki Takeda}
\affil{University of Miyazaki, Miyazaki, Miyazaki 889-2155, Japan}


\begin{abstract}%
We developed a beam telescope system comprising five layers of 300-$\mu$m-thick INTPIX4NA monolithic pixel sensors with each pixel size of 17 $\mu$m square. The sensors were fabricated using silicon-on-insulator (SOI) technology. The signal-to-noise ratio of 140--230 is realized at a bias voltage of 20~V.  
The tracker system was tested using a positron beam of 200--822 MeV/c, and various tracking methods are examined to optimize spatial precision achievable at these energies. The best tracking precision including the precision of the sensor under test itself is 11.04 $\pm$ 0.10 $\mu$m for 822-MeV/c positrons for an equidistant sensor spacing of 32 mm.
The achieved precision results combined with the intrinsic spatial resolution value obtained for a similar system using 120 GeV protons are used to estimate the tracking performance of electrons in the GeV energy range; a tracking precision of 2.22 $\mu$m is evaluated for 5-GeV electrons. The method to estimate the tracking performance is verified using a Geant4-based simulation.
The developed high precision tracker system enables to map the detailed performance of the sensors with pixel sizes of $\mathcal{O}$(10 $\mu$m), therefore will be a powerful system for development of devices targeting precision position resolutions.  

\end{abstract}

\subjectindex{H12}

\maketitle

\section{Introduction}

Precision beam tracking system is considered essential for detector development especially of such devices targeting a precise spatial resolution. High-energy beams available at CERN and Fermilab are preferred with regard to negligibly small multiple scattering of the beam;  however, accessibility is sometimes limited. In Japan, the Research Center for Electron Photon Science (ELPH), Tohoku University,  provides a positron beam with a maximum momentum of 822 MeV/c[1], and an electron beam of 1--5 GeV/c is under commissioning at AR-TB of High Energy Accelerator Research Organization (KEK) [2]. 

We have initiated a research and development of a precision beam telescope system that can be operated at the aforementioned energy or energy range. 
The present telescope system is considered  as a precision beam tracker usable, for example,  for detailed efficiency mapping of pixel sensors with pixel size of $\mathcal{O}$(10 $\mu$m). Therefore, tracking precision well below the pixel size is preferred for electrons of  5 GeV available at KEK AR-TB.  In general, the beam tracker comprising of multiple tracking layers with precise spatial resolution performs best for high-energy particles. However, due to multiple scattering of GeV energy electrons, the number of tracking layers may need to be optimized and tracking method needs to be studied in such conditions of limited number of tracking layers. The sensor with superior spatial resolution is to be adopted for detailed study of the tracking performance, whereas the spatial resolution requirement may be less critical in environments with multiple scattering.

For this purpose, silicon-on-insulator (SOI) semiconductor technology has been adopted for monolithic pixel sensor fabrication. Using the 0.2-$\mu$m fully-depleted SOI technology of Lapis Semiconductor, we have fabricated  an integration-type sensor, INTPIX4 [3][4], one of the large Lapis SOI sensors (sensitive area is 14.1 $\times$ 8.7 mm$^2$), with each pixel size of 17 $\mu$m square.
The telescope comprises five layers of 300 $\mu$m thick INTPIX4NA, a revised version of INTPIX4. 

The telescope system was tested at ELPH with the positron beam at 200--822 MeV/c. In order to  evaluate the beam tracker position resolution, we regard one of the five sensors as a device under test (DUT). The DUT hit position and the intersection of the track reconstructed using the rest of the sensors are compared to evaluate the performance of the telescope. The track reconstruction is examined considering various combinations of the tracking layers.  
Moreover, the tracking performance with inserting a 2~mm-thick-aluminum plate is studied. This is to simulate the tracker performance degradation in testing other types of  DUT  -- thicker or with additional material such as readout board or cable.

The tracking residuals obtained for the beam momentum of 200--822 MeV/c combined with the previous data obtained using protons of 120 GeV [5] are used to estimate the tracking performance in the 1--5 GeV range at the KEK AR-TB. A simulation based  on Geant4 is also conducted to examine the estimated results.

\section{The beam telescope}

\subsection{INTPIX4NA sensor and SEABAS2 readout system}

The series of integration-type pixel sensors, named INTPIX, is one of the SOI sensors that have been actively developed. The main parameters of INTPIX4, one of the INTPIX series sensors, are listed in Table~\ref{table_INT4} [3]. The INTPIX4NA sensors, used for the telescope at the ELPH beamline, have the output buffers modified for faster response. Apart from the modification and the wafer (resistivity and thicknesses), the two sensors are identical. 

\begin{table}[!h]
\caption{Main parameters of the INTPIX4 and INTPIX4NA sensors.}
\label{table_INT4}
\centering
\begin{tabular}{|c|c|}
\hline
Chip size & 15.4 $\times$ 10.2 mm$^2$\\ 
\hline
Active area & 14.1 $\times$ 8.7 mm$^2$\\ 
\hline
Pixel size & 17 $\times$ 17 $\mu$m$^2$\\ 
\hline
Pixel array & 512 (row) $\times$ 832 (column)\\ 
& 13 blocks (64 cols) in parallel readout\\
\hline
Wafer& $\rho \sim$7~k$\Omega$cm, 500~$\mu$m thick  (INTPIX4)\\
  n-type FZ    & $\rho \sim$11~k$\Omega$cm, 300~$\mu$m thick  (INTPIX4NA)\\
\hline
modification& Output buffer is enforced for NA\\
 
\hline
\end{tabular}
\end{table}

INTPIX4 (and INTPIX4NA) employs a correlated differential sampling method to suppress the noise associated with switching. The charge integrated for predetermined duration (9~$\mu$s at ELPH testbeam) is reset periodically unless the STORE signal is received. The STORE triggers readout sequence, holding the charges stored in the capacitors and sending them one-by-one to external 12-bit analog-to-digital converters (ADCs). 

The readout of each INTPIX4NA sensor is controlled by the SEABAS2 [4] readout system.
The INTPIX4NA chip is glued to a separate board, i.e., a sub-board, which is connected to the SEABAS2 board via a pair of 64-pin connectors.
The groups of 64-column data (times 512 row data) are fed into one ADC; overall, 13 ADCs are operating simultaneously   on the SEABAS2 board.  The digitization time per pixel (or scan time), including switching time to the next pixel data is set to 200~ns at the testbeam, whereas it functions reliably at a scan time of 150~ns.
SEABAS2 is equipped with two 
field-programmable gate arrays (FPGAs), one (Virtex-5) for user programming and the other (Virtex-4) for data transfer with 1~GbE SiTCP protocol.

\subsection{The telescope system evaluated with the ELPH positron beam}

\begin{figure}[!h]
\centering\includegraphics[width=5in]{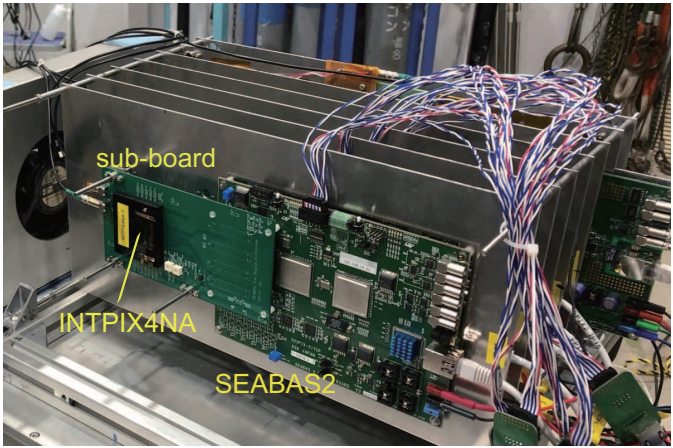}
\caption{The telescope system placed in the ELPH beam, viewed from beam upstream. There are five INTPIX4NA sensors, followed by one XRPIX5 sensor. The supporting aluminum boards are 20 cm high and 40 cm wide. The sensor is located behind the black block (the beam window at center is taped in this photo) on the sub-board, which is connected to the SEABAS2 board. The control signals (RESET, STORE, TimeStamp Clock, BUSY) are transmitted through the 10-pin connector at the top of SEABAS2 board, with the ADC data extracted through the RJ-45 connector at right.}
\label{fig_setup}
\end{figure}

\begin{figure}[!h]
\centering\includegraphics[width=4in]{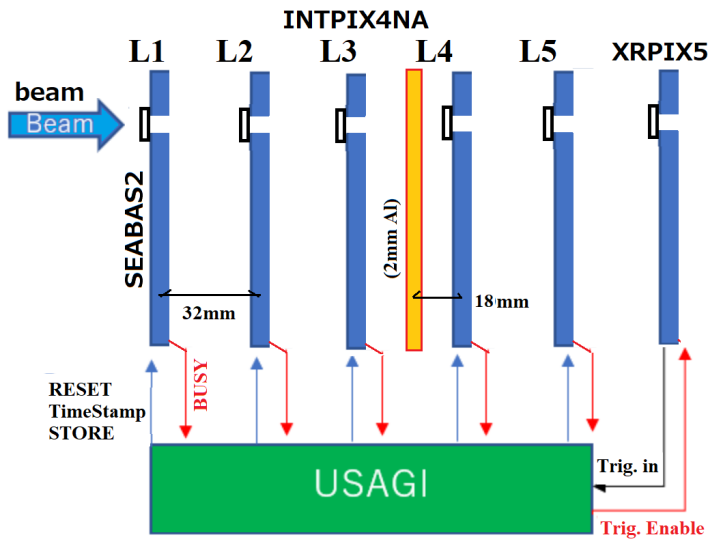}
\caption{Five telescope layers followed by one XRPIX5 sensor, with the trigger logic unit (USAGI). A 2-mm thick aluminum plate, shown in orange, was placed later in the measurement to mimic the effect of additional material.}
\label{fig_setup2}
\end{figure}

The telescope system comprises five INTPIX4NA sensors and another beam-triggering SOI pixel sensor, XRPIX5 [6], located at the downmost-stream position. Both INTPIX4NA and XRPIX5 are equipped with a function defining the region-of-interest (ROI).
 The INTPIX4NA sensors are named L1 to L5 counting from upstream.
A photograph of the complete structure of the telescope is shown in Figure~\ref{fig_setup}.
The sensors are placed at an equi-distance of 32~mm (Figure~\ref{fig_setup2}). 
The SEABAS2 board is screwed to a 2-mm thick aluminum board whose height is 20~cm and width is 40~cm. Six aluminum boards are subsequently screwed using 30-mm long spacers. The sub-boards of the five sensors are pre-aligned using three stainless rods (visible in Figure~\ref{fig_setup}) at a precision of 5~$\mu$m. 

The data acquisition sequence is controlled by a trigger logic unit called USAGI equipped with an FPGA. USAGI distributes  synchronized RESET signals at a 100~kHz frequency and 1~kHz timestamp signals to the five INTPIX4NA sensors. On receiving the beam trigger signal from XRPIX5, USAGI disables XRPIX5 from generating further triggers and sends the STORE signal. Then, INTPIX4NAs terminate generating further RESET signals and the readout sequence is initiated individually in each SEABAS2 board. The BUSY signal is generated by each SEABAS2 board while data digitization is in progress and while RESET is set. The USAGI receives BUSY signals from the SEABAS2 boards and clears the XRPIX5 disable signal once all the BUSY signals are cleared. The timestamp count is attached to the data to verify the simultaneity of the hits. The inactive INTPIX4NA sensors while BUSY is set are not exactly synchronized among the five layers; however more than 99\% of the data have been found to be synchronized.

The telescope system possesses a beam window, and only sensors with a thickness of 300~$\mu$m are exposed to the beam. Later, we placed a 2-mm thick aluminum plate before L4, as illustrated in Figure~\ref{fig_setup2}, to mimic additional materials and investigate the effect of the materials on beam tracking performance.

The ELPH provides positron beams with a maximum momentum of 822~MeV/c. Photons from the ELPH 1.3 GeV electron synchrotron are converted on a 200-$\mu$m thick tungsten target, and positrons with specific momentum are selected  with a relative spread of 0.85\% for a beam of 822 MeV/c. Most measurements were performed at this momentum. In addition, low momenta, down to 200~MeV/c, were considered to examine the momentum dependence. The relative momentum spreads are 2.5\% at 200 MeV/c, 1.8\% at 300 MeV/c, and 1.2\% at 500 MeV/c [7].

The positron beam was extracted for $\sim$10~s followed by a quiet time of 7~s. The beam was transmitted through a set of quadrupole magnets with the resulting beam profile at the telescope being wide enough, such that the beam distributions were almost flat along the horizontal and vertical directions.

\section{Performance evaluation using the ELPH positron beam}
\subsection{Cluster charge and signal-to-noise ratio}

The pedestal of an individual pixel is calculated by averaging the ADC values obtained in the beam data.  Simple averages resulted in the pedestal rms spreads \sped of 4--8 ADCs fluctuating among data taking runs.  The main contribution to the spreads was identified to be due to events where pedestals of all pixels shifted by an almost constant amount, which occurred intermittently. To correct for such pedestal jump,  the average of all pixel pedestals was calculated on an event-by-event basis, which was used for the pedestal correction. The average calculation was made for entire pixels deriving 1 common correction value or for each of the 13 readout blocks deriving 13 correction values. These corrections reduced the pedestal fluctuation to \sped of 1.9--2.8 ADCs with a single value or 1.6--2.2 ADCs with 13 value corrections. We adopted the pedestals with per block frame correction (13 value corrections) in the following analysis.     

\begin{figure}[!h]
\centering\includegraphics[width=6in]{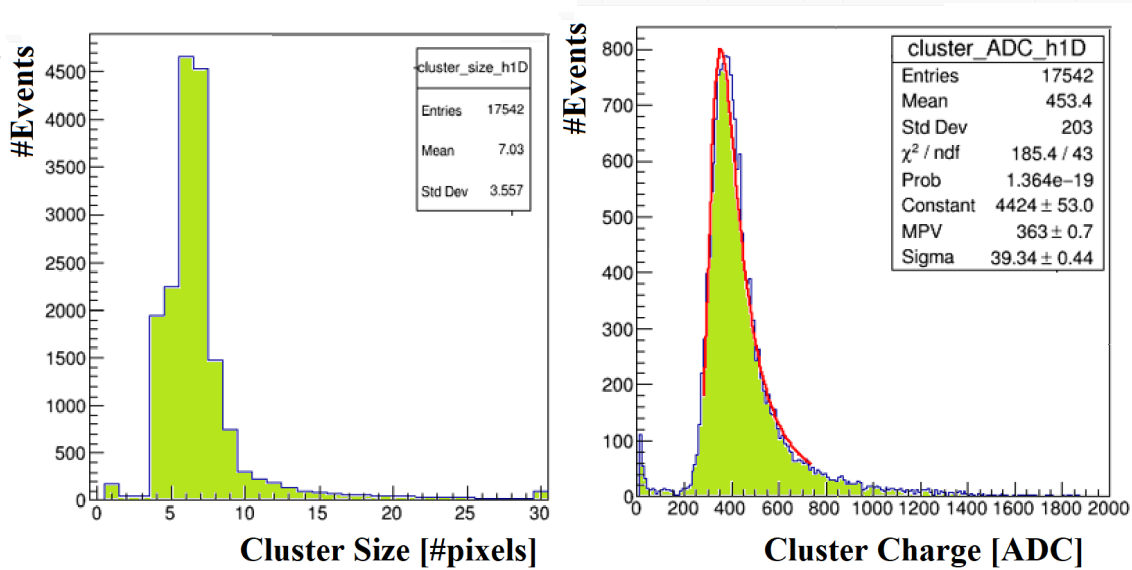}
\caption{The distribution of number of pixels in a cluster (left) and of the cluster charge (right) of L1 INTPIX4NA sensor at 20~V bias voltage.}
\label{fig_CS}
\end{figure}

Clustering is a step to correlate the pixel data to the beam hit. We first find the cluster-seed pixel which has the largest signal and with more than 10$\sigma$ of pedestal (typically more than 20 ADC counts).  The pixels with more than 10$\sigma$ of pedestal about the seed pixel are added to construct a cluster. Figure~\ref{fig_CS} shows the distribution of cluster sizes (CS) defined by the number of pixels included in a cluster, and cluster charges (CC) of the L1 sensor operated at 20~V bias. The cluster size distribution has a mean of 7.0 pixels extending up to 20 pixels. The CC exhibits a Landau distribution (overlaid in the plot) with a most probable value (MPV) of 363 ADCs. The events with small cluster charge are for the clusters of which seeds are located near the effective edge of the sensor. 

The MPVs of CC distributions and pedestal sigmas (\sped) obtained at 20~V are listed in Table~\ref{table_SN}, along with the signal-to-noise (S/N) ratios defined as MPV divided by \sped. The S/N ratios are notably large, as have been achievable with SOI pixel sensors [8]. 

\begin{table}[!ht]
\caption{S/N ratios at 20~V bias for the five layers. Noise and signal MPV values are defined in the unit of ADCs.}
\label{table_SN}
\centering
\begin{tabular}{|c|ccc|}
\hline
layer & noise \sped & signal MPV &S/N\\ 
\hline \hline
L1 & 1.6 & 363 & 225 \\ \hline
L2 & 1.8 & 327 & 181 \\ \hline
L3 & 1.8 & 341 & 186 \\ \hline
L4 & 1.7 & 262 & 148 \\ \hline
L5 & 1.6 & 364 & 230 \\ \hline 

\end{tabular}
\end{table}

\subsection{HV dependence of cluster characteristics}

\begin{figure}[!h]
\centering\includegraphics[width=4in]{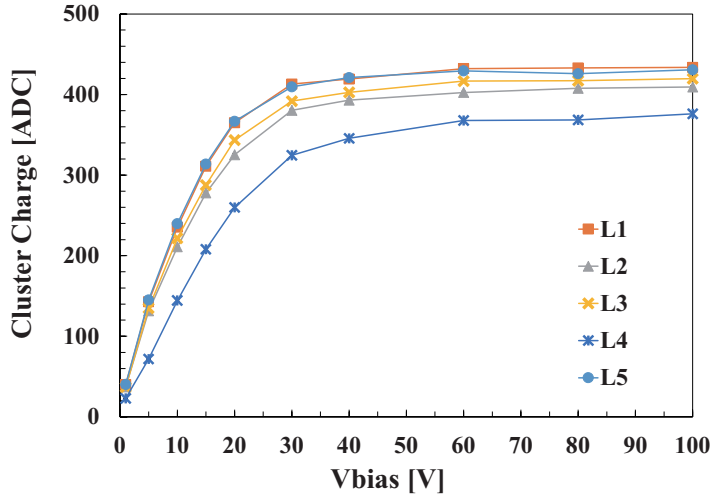}
\caption{The most probable CC values as a function of  bias voltage shown for five INTPIX4NA sensors.}
\label{fig_CCV}
\end{figure}

The bias voltage dependence of the CCs is plotted in Figure~\ref{fig_CCV}.  The curve exhibits a plateau above $\sim$30~V, implying that the sensor is fully depleted. The full depletion voltage defined by the voltage of the intersection where the plateau constant and linear line fitted to CC vs. \Vbias data points with \Vbias $\leq$ 20~V ranges from 26~V (L1 and L5) up to 38~V (L4). As the nominal wafer resistivity is in the range of 10.1--12.1 k$\Omega$cm, the expected full depletion voltage is in the range of 27--33~V for 
300-$\mu$m-thick $n$-type silicon; the estimation is almost consistent with the measured voltages. 

\begin{figure}[!h]
\centering\includegraphics[width=4in]{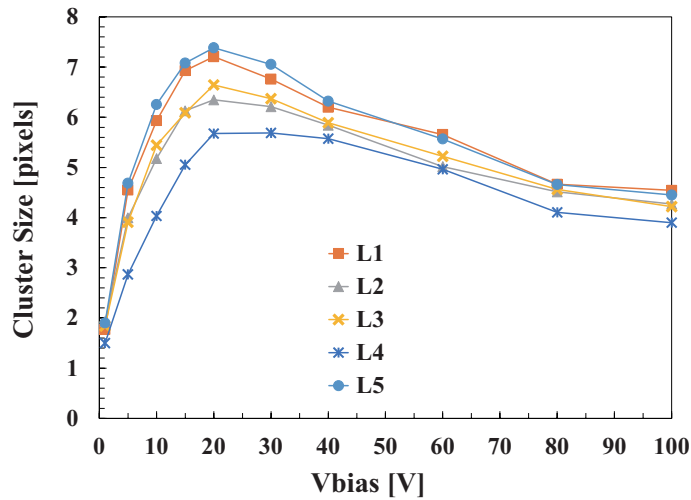}
\caption{The average number of pixels in a cluster (cluster size) as a function of  bias voltage shown for five INTPIX4NA sensors.}
\label{fig_CSV}
\end{figure}

The bias voltage dependence of the CSs is plotted in Figure~\ref{fig_CSV}.  As an adequate charge spread with regard to the hit position sharing the charge among multiple pixels  is a key for superior spatial resolution, the curves are characteristic; the CS increases with the bias below the full depletion level as the carrier drift length increases and then CS decreases above the full depletion as the electric field increases therefore diffusion effect is suppressed.

We observe slight variations in the S/N, CC, and CS distributions among the five layers. This is explained by the individual variation in  INTPIX4NA gain settings [3] in the SEABAS2 boards, which is verified after the beam test was completed. However, the bias dependence curves are almost similar among the five layers.

\subsection{Tracking precision for 822 MeV/c positrons}

The hit position is defined by the ADC-averaged weight mean of the hit pixels' positions. Tracks are defined by the hits of a certain combination of layers excluding the DUT.  The tracking precision is defined as the standard deviation of the hit residuals of DUT with respect to the track crossing point at the DUT.  A study of the tracking precision is performed by changing the DUT and combinations of tracking layers. The track hit position after completing the detector alignment is denoted as $X$ and $Y$ on the sensor face with $Z$ along the beam direction.

In high-energy hadron beams available at CERN and Fermilab, the detector intrinsic spatial resolution is obtained by fitting the detector hits to a straight line as the effects of multiple scattering is negligibly small.   
The intrinsic resolution of the INTPIX4 derived is 1.34--1.65~$\mu$m [6], as described in Section~\ref{sec_fnal}.

\begin{figure}[!h]
\centering\includegraphics[width=6in]{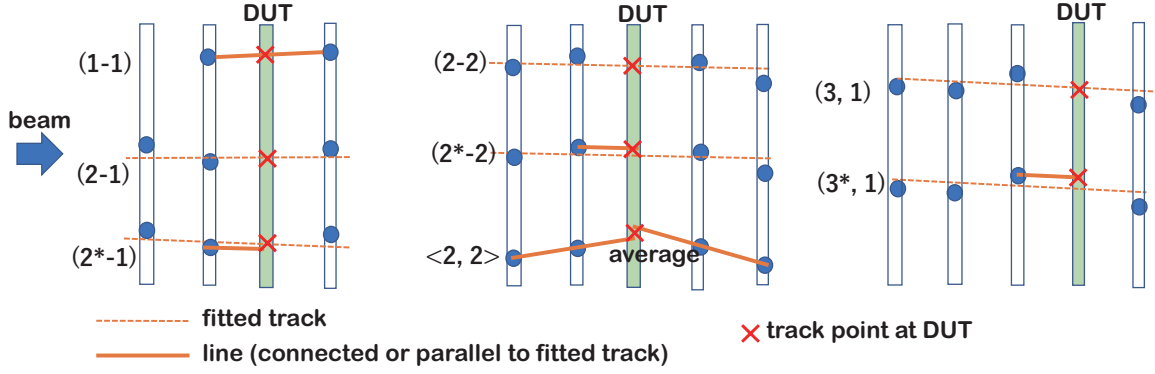}
\caption{Illustration of the investigated tracking methods. The track point on the DUT is calculated as \color{red}$\times$\color{black} \, using the hit information (marked as \color{blue} $\bullet$\color{black}) from other layers. For the cases of 3 and 4 tracking layers, the hit points are fitted to a straight line. The methods (2*-1), (2*-2) and (3*-1) use the hit nearest to the DUT and a line is extrapolated to the DUT using the slope of the fitted line.
 The illustrations are for the beam directed as shown by the arrow. Other methods, such as  (1-2) and (1-*2),  correspond to the cases where the beam is directed towards the left.
 The method $<$2,2$>$ takes the average of  the two lines at the DUT, as discussed in Sec.~\ref{material}. }
\label{fig_tracking}
\end{figure}

For the electrons with 822 MeV/c, the tracking has to be optimized considering the multiple scattering. As illustrated in Figure~\ref{fig_tracking}, first, we consider Method (1-1) where one layer in each side of the DUT is used for tracking.
Secondly, in Methods (2-1) to (1-*2),  two layers on one side and one layer on the other side of the DUT are used. In these cases, three layer hits are fitted to a straight line. The intersection at the DUT is calculated in Methods (2-1) and (1-2)  using the fitted track. In Method (2*-1), the fitted track slope is used but the track is extrapolated from the hit recorded just in front of the DUT, and similarly in Method (1-*2) the track is extrapolated from the hit recorded just at the rear of DUT. The results are listed in Table~\ref{table_res}.  The values vary slightly depending on the selection of the DUT and X/Y directions, which should be explained by the differences in the individual detector  performance and mostly by the residual miss-alignments. Averages of the values are calculated per method, and shown in the table. The statistical uncertainty or maximum variation of the central values are taken as the uncertainty.
Methods (2-2) and (2*-2) use the two planes each side of the DUT, and Methods (3-1) to (1-3) use three planes on one side and one plane on the other side of the DUT. In these cases, the track is fitted using the selected planes and in the methods with * the track is extrapolated from the hit closest to the DUT.

\begin{table}[!ht]
\caption{Residual sigmas [$\mu$m] for 822-MeV/c positrons measured at 20~V bias. Two values in each field are measured along the $X$ and $Y$ directions.}
\label{table_res}
\centering
\begin{small}
\begin{tabular}{|c|ccc|c|}
\hline
Tracking &  & DUT &  &Average\\ 
method & L2 & L3 & L4 &\\ 
\hline \hline
(1-1)  &12.06$\pm$0.10/11.91$\pm$0.10 &  12.15$\pm$0.10/12.24$\pm$0.10 &12.71$\pm$0.12/12.98$\pm$0.11&  12.34$\pm$0.11 \\ \hline 
(2-1)  & - &14.42$\pm$0.12/14.54$\pm$0.12&14.91$\pm$0.13/15.33$\pm$0.13&14.80$\pm$0.12\\ \hline
(2*-1) & - &11.80$\pm$0.10/11.72$\pm$0.12&12.23$\pm$0.10/12.40$\pm$0.10&12.04$\pm$0.11\\ \hline 
(1-2)  &14.32$\pm$0.12/14.27$\pm$0.17&14.48$\pm$0.12/14.48$\pm$0.12& - &14.39$\pm$0.12\\ \hline
(1-*2) &11.72$\pm$0.10/11.63$\pm$0.10&12.18$\pm$0.10/12.22$\pm$0.12& - &11.94$\pm$0.11\\ \hline 
(2-2) & - &20.28$\pm$0.16/20.54$\pm$0.17& - &20.31$\pm$0.17\\ \hline 
(2*-2) & - &14.57$\pm$0.12/14.63$\pm$0.17& - &14.60$\pm$0.15\\ \hline 
(3-1) & - & - & 16.28$\pm$0.14/16.11$\pm$0.14 &16.20$\pm$0.14\\ \hline 
(3*-1) & - & - & 15.27$\pm$0.12/14.27$\pm$0.17 &14.77$\pm$0.15\\ \hline 
(1-3) & 14.32$\pm$0.12/14.27$\pm$0.17 & - & - &14.30$\pm$0.15\\ \hline 
(1-*3) & 14.87$\pm$0.14/15.02$\pm$0.12 & - & - &14.95$\pm$0.13\\ \hline 
$<$2*-1,1-*2$>$& - &11.10$\pm$0.10/10.98$\pm$0.09& - &11.04$\pm$0.10\\ \hline 
\end{tabular}
\end{small}
\end{table}

Method (1-1) provides an optimal value of 12.34$\pm$0.11 $\mu$m, and any additional hit information degrades the tracking precision, unless the hit nearest to the DUT is used as an anchor as in Methods (2*-1) and (1-*2), where slightly better residual of 12 $\mu$m has been obtained. The best performance 11~$\mu$m is obtained for Method $<$2*-1,1-*2$>$ where the tracker positions obtained by Methods (2*-1) and (1-*2) are averaged. The intrinsic INTPIX4 resolution is included in these values; however, the value of $\sim$1.5~$\mu$m is considerably smaller and not subtracted.   

In general, best tracker resolution is achievable for higher-energy particles of which trajectories are reconstructed using multiple sensor planes with finer pixel pitch.
For lower energy electrons, due to multiple scattering,  requirements of fine pitch pixels and multiple layers may become less important. However, as shown in the study presented here, while more than two planes on one side are not necessary, sensors with fine pixel pitch play an important role to set the track anchor point near the DUT.    

\begin{figure}[!h]
\centering\includegraphics[width=4in]{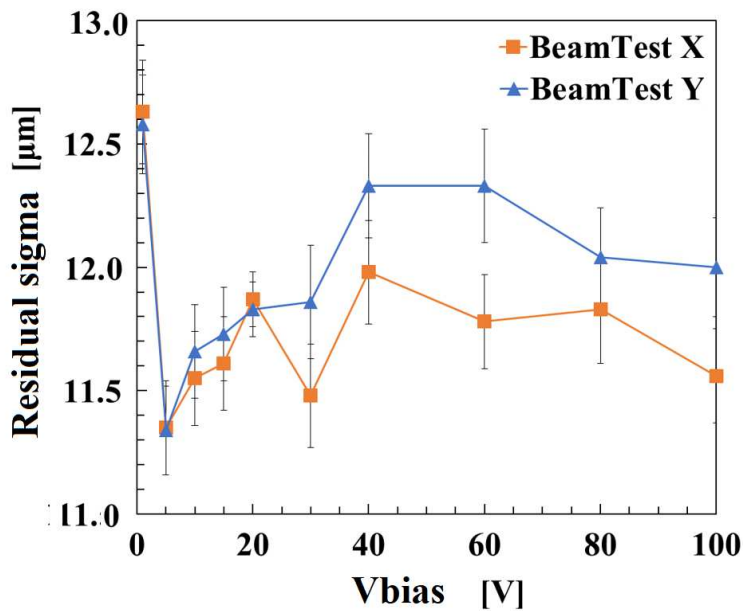}
\caption{The residual sigma of (2*-1) tracks with respect to  L3 hits as a function of  bias voltage, demonstrated in $X$ and $Y$ directions.}
\label{fig_ResV}
\end{figure}

The bias dependence of (2*-1) tracking precision with L3 chosen as the DUT is plotted in Figure~\ref{fig_ResV}.
The dependence is weak, where the spatial resolution improves with a decrease in the bias except at 2~V. At 2~V, the S/N ratio is roughly 1/10 of that at 20~V, and the charge spread is not large enough.

As a good spatial resolution is realized by adequate charge spread among pixels with good S/N, the use of the SOI sensor that can realize fine pixel size and superior S/N is the key for the high-performance tracker. For example in [8], we demonstrated that the intrinsic position resolution of SOI pixel sensor FPIX with 8-$\mu$m pixel size and S/N of $\sim$300 is 0.60--0.75 $\mu$m, the first semiconductor device with sub-micron spatial resolution.

\subsection{Tracking with additional material}\label{material}

\begin{figure}[!h]
\centering\includegraphics[width=3.5in]{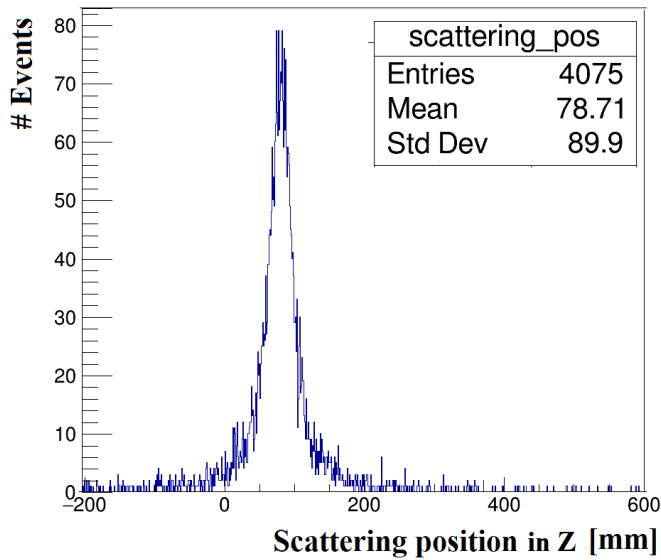}
\caption{The scattering position in  $Z$ given by the closest approach point of L1--L2 and L4--L5 tracks.}
\label{fig_scatteringZ}
\end{figure}

Tracking needs to be optimized depending on the thickness of the DUT. In the ELPH beam line, we added a 2-mm-thick aluminum plate between L3 and L4 (Figure~\ref{fig_setup2}) to simulate the performance degradation of the present setup for the cases with thicker DUT. 
One of the robust tracking methods against material, as far as the beam track passes through the DUT, is to use two separate tracks reconstructed using the layers in front and those at rear.
 Figure~\ref{fig_scatteringZ} plots the distribution of the position along the beam defined as the intersection of  the track given by L1 and L2, and that given by L4 and L5. The mean value of 78.7 mm with regard to  L1 is consistent considering the location of the aluminum plate. The average of the extrapolation points at the DUT of the two tracks is considered and denoted as Method $<$2,2$>$.

\begin{table}[!ht]
\caption{Residual sigmas [$\mu$m] for 822 MeV/c positrons measured at 20~V bias. The values are for  DUT = L3 compared with and without the 2-mm-thick aluminum. Method $<$2,2$>$ refers to the average position of the tracks reconstructed separately in front and at the rear of the DUT, see Figure~\ref{fig_tracking}.}
\label{table_Al}
\centering
\begin{tabular}{|c|cc|}
\hline
Tracking method & without Al & with Al  \\ 
\hline \hline
(1-1)  &12.28$\pm$0.10   & 17.92$\pm$0.24 \\ \hline 
(2*-1)  &11.78$\pm$0.12   & 15.27$\pm$0.21 \\ \hline 
(1-*2)  &12.20$\pm$0.12   & 13.57$\pm$0.20 \\ \hline 
$<$2*-1,1-*2$>$  &11.04$\pm$0.10   & 11.60$\pm$0.19 \\ \hline 
$<$2,2$>$  &17.55$\pm$0.17   & 18.76$\pm$0.26 \\ \hline 
\end{tabular}
\end{table}

Table~\ref{table_Al} lists the effect of 2-mm-thick aluminum plate on the tracking residual sigmas for the tracking methods that showed good performance in Table~\ref{table_res}. Methods (1-1) and (2*-1) are significantly degraded by the additional material. The degradation is modest for Method (1-*2) as the method can trace the scattering at the Al located at the rear of DUT, compared to Method (2*-1). It is notable that Method $<$2*-1,1-*2$>$  maintains the best performance for both cases. Method $<$2,2$>$ also exhibits  relatively small degradation though the initial value is larger than the other methods.   

The additional material of 2-mm Al is considered to be thick enough for a tracker DUT considering additional readout board/cable exposed in the beam. As Method $<$2*-1,1-*2$>$ shows best performance, the present telescope is verified to work for such cases.   
However,  in the case of very thick DUT, like a calorimeter, we need to reconstruct the track using the layers in front only.
We obtain 24.46 $\pm$ 0.21 $\mu$m for (2-0),  0 meaning that no layer at rear is used,  and  27.82 $\pm$ 0.23 $\mu$m for (3*-0). We observe that the third layer does not improve the resolution for 822 MeV/c electrons as multiple scattering is significant. 

\section{Tracking precision expectation for electrons within the GeV energy range}

\subsection{Intrinsic resolution evaluation with 120-GeV protons}~\label{sec_fnal}
We evaluated the intrinsic INTPIX4 resolution [6] using 120-GeV protons at the Fermilab Test Beam Facility in 2018. The tracker system comprises seven layers in total with three DUTs (SOFIST [9], being developed for the ILC [10] experiment) sandwiched by two tracker INTPIX4 planes in each side. The sensor thickness of INTPIX4 was 500~$\mu$m instead of 300~$\mu$m of INTPIX4NA. The different sensor thickness can be ignored as multiple scattering is negligible and CS distributions were similar to those obtained at the ELPH beam test; the mean values  ranged from 7.2 to 9.3 pixels for  the four INTPIX4 sensors.
 
\begin{figure}[!h]
\centering\includegraphics[width=4in]{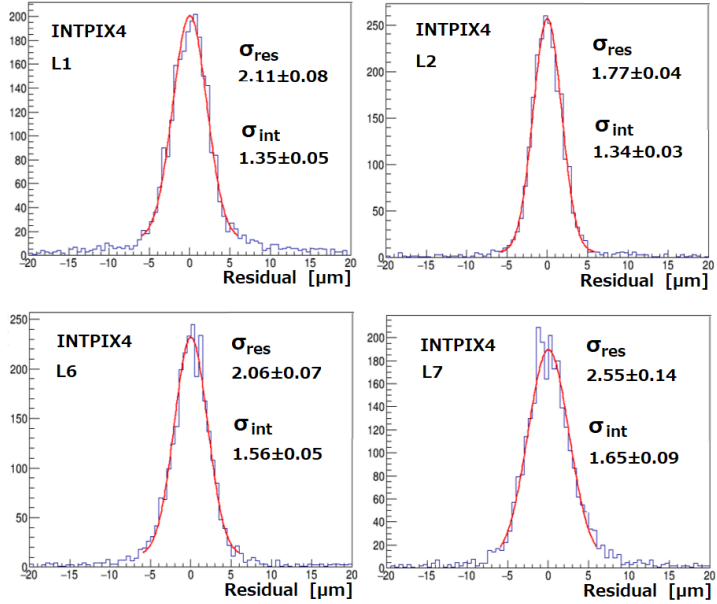}
\caption{The tracker residual distributions of four INTPIX4 sensors measured using 120-GeV protons. $\sigma_{\rm res}$ is of the fitted Gaussian, and   $\sigma_{\rm int}$ is the intrinsic sensor resolution corrected for the tracker uncertainty.  The unit is in $\mu$m.}
\label{fig_120GeV}
\end{figure}

At the equidistant spacing of 32~mm,  three SOFIST sensors were set at the middle with the four  INTPIX4 sensors set and numbered as L1, L2, L6 and L7. Figure~\ref{fig_120GeV} shows the residual of the sensor hits to the tracks reconstructed using the other three INTPIX4 sensors.  The Gaussian sigmas ranges from 1.77 to 2.55~$\mu$m.  After subtracting the tracker resolution, the intrinsic resolution ranges from 1.34 $\pm$ 0.05 $\mu$m (for L2) to 1.65 $\pm$ 0.09~$\mu$m (for L7). 

 The average value of the obtained resolutions 1.50 $\pm$ 0.15 $\mu$m includes the residual miss-alignment but is taken as a conservative value of the intrinsic spatial resolution of INTPIX4 and INTPIX4NA, where the uncertainty is the maximum deviation of the four estimated values.

\subsection{Estimation of tracking precision in GeV energy range}

\begin{figure}[!h]
\centering\includegraphics[width=3.5in]{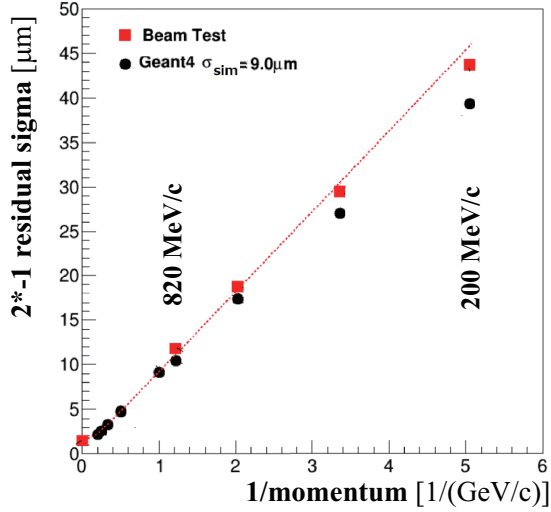}
\caption{The residual sigma of (2*-1) tracks with respect to the L3 hits as a function of  1/momentum.}
\label{fig_mom}
\end{figure}

 The intrinsic resolution of 1.50 $\pm$ 0.15 $\mu$m corresponds to the (2-1) track residual sigma of $1.84 \pm 0.18$ $\mu$m at 120 GeV, where the uncertainty originates from the uncertainty of INTPIX4 intrinsic resolution described above. 
The obtained value and four (2*-1) residual sigmas measured at the four momenta at ELPH are fitted to  the following function;
\begin{equation}\label{1}
\begin{split}
\sigma_{\rm res}(p) = \sqrt{\sigma_0^2 + (k/p)^2},
\end{split}
\end{equation}
\noindent
with resulting value  $k=9.023 \pm 0.077$ $\mu$m$\cdot$GeV/c, and  $\sigma_0 =1.84 \pm 0.18$ $\mu$m from the Fermilab testbeam result.
The data points and fitted function are plotted in Figure~\ref{fig_mom}.

Also plotted in the figure are the evaluations at ELPH momenta and 1-5 GeV electrons based on a Geant4  [11] simulation.
While Geant4 precisely traces the electron trajectory considering detailed physical interactions with the material, the conversion from the energy deposition in the sensor to the pixel charge spread needs to be implemented additionally.  
Instead of relying on dedicated simulation such as TCAD, the beam data are used to parameterize the charge spread.

\begin{figure}[!h]
\centering\includegraphics[width=3.5in]{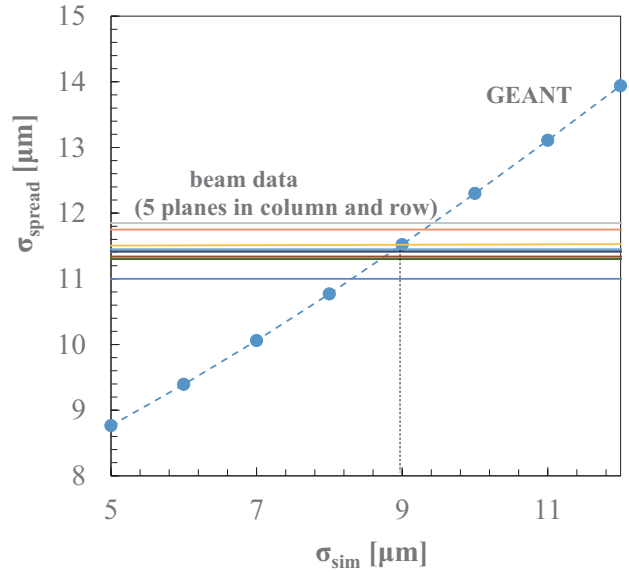}
\caption{The relation between the Gaussian charge spread $\sigma_{\rm sim}$ implemented in the simulation and the pixel charge spread $\sigma_{\rm spread}$ about the seed pixel.   $\sigma_{\rm spread}$ values obtained in the beam data for all five layers both in column and row  directions are shown in the ordinate axis.}
\label{fig_geant}
\end{figure}

The distribution of pixel charges normalized by the maximum pixel charge in an event is projected in pixel column and row directions, and then, the projected distributions are fitted with a Gaussian function. Figure~\ref{fig_geant} shows the Gaussian sigmas $\sigma_{\rm spread}$ on the ordinate axis obtained in the ELPH beam for the five layers in both  column and row directions, in total 10 values. The value ranges from 11.0 to 11.8 $\mu$m. In the Geant4 simulation, the deposited energy is distributed in the column and row directions according to  Gaussain with a given spread $\sigma_{\rm sim}$, and the pixel charges are defined as the integral of the charge in the corresponding pixel area of 17$\times$17~$\mu$m$^2$. The electron beam is injected uniformly across the pixel area. With an increase in the $\sigma_{\rm sim}$ value,  $\sigma_{\rm spread}$ obtained with regard to the beam data increases as shown in the figure. The optimum $\sigma_{\rm sim}$ is set to 9~$\mu$m from this plot. Further steps are added in the simulation to correct for the charge spread in the pixels apart from the seed pixel and for the finite ADC pedestal spread.

The (2*-1) track residual sigma estimated for 5-GeV electrons is  $2.68 \pm 0.14$ $\mu$m from Equation (1) and  $2.21 \pm 0.02$ $\mu$m  based on the Geant4 simulation. The difference in the estimation should originate in (I), where the beam data overestimate the resolution as there is residual layer misalignment, and (II), where the simulation does not consider for event-by-event fluctuation in the charge spread nor position dependent charge spread, generating an underestimated value.     
Although the simulation does not fully describe the measured sigmas especially in the sub-GeV energy range, the agreement between the simulation and Equation (1) estimations in the GeV energy range is acceptable.

\section{Comparison with the existing tracker and possible system improvements}

The current tracker system has been constructed using the available INTPIX4NA sensors and we have evaluated the spatial resolution achievable with this system for sub-GeV to GeV electrons.  Comparison with other existing tracker is interesting. We discuss remaining issues on possible improvements of the telescope performance. 

\subsection{Comparison with the EUDET tracker}
One of the performance beam trackers suitable for electron energy ranging from sub-GeV to GeV is EUDET beam telescope [12] based on MIMOSA26 monolithic active pixel devices.
The MIMOSA26 sensor consists of pixels sized 18.4 $\times$ 18.4 $\mu$m$^2$ and a sensor thickness of 54~$\mu$m.
For 6 GeV electron/positron beam, the spread of the biased residual distribution, including tracker and DUT resolutions, was 2.88 $\pm$ 0.08~$\mu$m. By subtracting the DUT intrinsic resolution of 3.24 $\pm$ 0.09 $\mu$m, the track resolution using an equidistant plane spacing of 20~mm is estimated to be 1.83 $\pm$ 0.03 $\mu$m for 5-GeV electrons.

By subtracting the INTPIX4NA intrinsic resolution of 1.5~$\mu$m, the tracking residual sigma values mentioned in the previous section correspond to 2.22 $\mu$m of the beam data interpolation and 1.62~$\mu$m of the Geant4-based simulation; the values are comparable to the EUDET evaluation of 1.83 $\pm$ 0.03 $\mu$m. 

\subsection{DAQ rate and possible improvement}

\begin{figure}[!h]
\centering\includegraphics[width=4in]{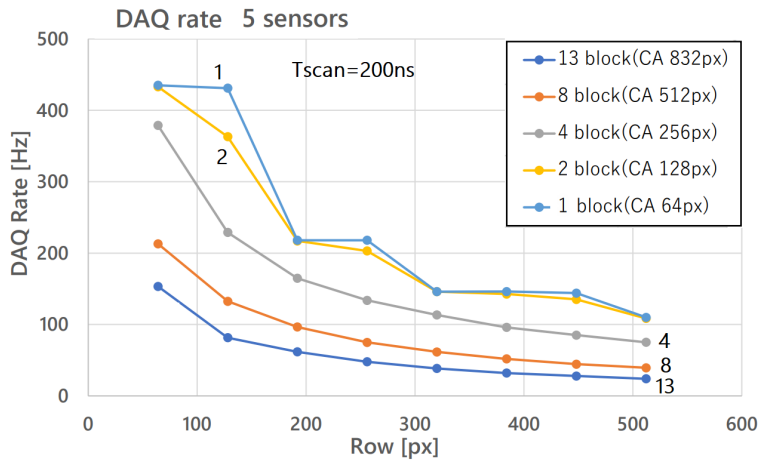}
\caption{The DAQ rate of the 5-layer system as a function of the number of readout row pixels and for the number of readout column blocks (each block contains 64 columns). The pixel scan scan time is set to 200~ns.}
\label{fig_DAQrate}
\end{figure}

The present tracker system employs a switching hub to collect all the five layers of INTPIX4NA data and one layer of the  XRPIX5 data to a computer. The DAQ rate when all the five layers are read out is plotted in Figure~\ref{fig_DAQrate} as a function of the number of readout row pixels (up to 512) for various settings of the number of readout column blocks (up to 13). The DAQ rate of single sensor is primarily determined based on  the pixel scan time, which is 200~ns in this data sample. As the system can be operated at 150~ns, the rate will increase accordingly. The scan time as a 5-layer system is degraded further with an increase in the number of readout blocks, thereby implying that the data transfer rate to the computer limits the overall rate. By reducing the amount of transfer data by employing zero-suppression in the FPGA, the rate should recover to 105 Hz. 

The maximum beam rate expected at the KEK AR-TB is about 800 Hz in the sensor area, therefore further improvement of readout speed is preferred. 
As described before, the analog signals generated from INTPIX4NA are digitized using 12-bit ADCs located in the SEABAS2 board. By placing ADCs near to the INPIX4NA chip on the sub-board should help reducing the scan time. The simulation suggests that a scan time of 100~ns should be achievable. Reducing the ADC bits shortens the digitization time.   Research on these issues are in progress.
Ultimate improvement is implementation of column ADCs as successfully implemented in the SOFIST [10] sensor, which is beyond the scope of our development.

\subsection{Improvement with thinner sensors}

\begin{figure}[!h]
\centering\includegraphics[width=4in]{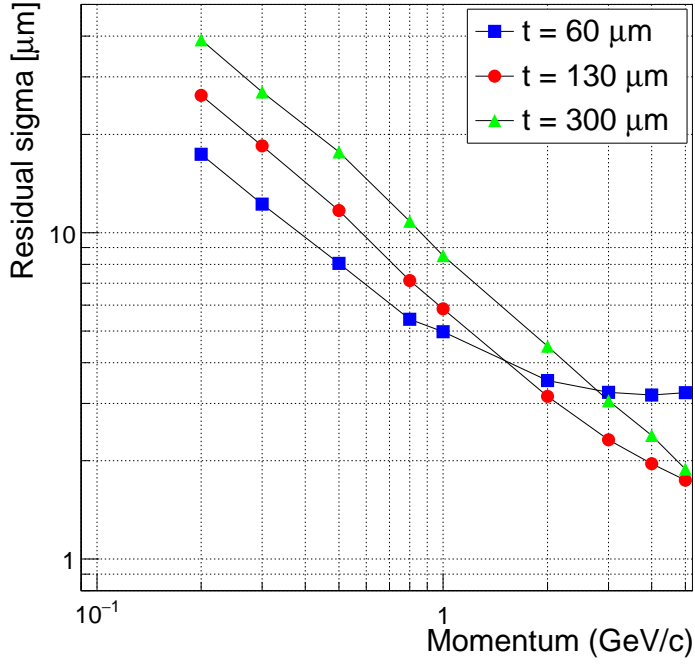}
\caption{(2*-1) track residual sigmas as a function of electron momentum estimated using Geant4 based simulation for the sensor thicknesses of 60, 130 and 300 $\mu$m.}
\label{fig_thickness}
\end{figure}

The bias voltage dependence of residual sigma shown in Figure~\ref{fig_ResV} ensures that the tracking performance does not degrade much at a bias of 5~V, where the average CC is 140 ADCs (40\% of the cluster charge at 20~V). To improve the tracking performance of low-energy electrons using thinner sensors, we have estimated the performance when the sensors are 60~$\mu$m or 130~$\mu$m thick using the same Geant4-based simulation with the same setting of  
$\sigma_{\rm sim} = 9~\mu$m. The simulation results are illustrated in Figure~\ref{fig_thickness}, where (2*-1) track residual sigma values are plotted as a function of electron momentum. 
 In this range, the tracker system with 130-$\mu$m thick sensors exhibits the best performance, whereas the system with 300-$\mu$m sensors degrades more at lower momenta as expected. The performance of 60-$\mu$m thick sensors does not improve with momentum as the signal size is small, thereby generating low S/N.
The constructed system using 300-$\mu$m thick sensors similarly performs as 130-$\mu$m thick sensors for 5 GeV/c electrons.

\section{Summary}

We have constructed a beam tracker using INTPIX4NA SOI  pixel sensors and evaluated the tracking performance of electrons within sub-GeV energy range. Various tracking methods are examined, and the best residual sigma of 11.04 $\pm$ 0.10 $\mu$m is obtained for 822-MeV/c electrons, including the DUT intrinsic resolution of  1.50 $\pm$ 0.15 $\mu$m.  The performance does not degrade when also for the case when the 2-mm aluminum plate was added simulating a case where the DUT is thicker or with additional material.

The measured data combined with the data derived using 120-GeV protons are used to evaluate the performance of the tracker system at GeV electrons. The Geant4-based simulation describes the measured results reasonably well.   The tracker resolutions, including the intrinsic DUT resolution, are estimated to be 2.68~$\mu$m of the beam data and  2.21~$\mu$m of the Geant4-based simulation for 5 GeV electrons. 
The developed high precision tracker system enables to map the performance of the sensors with pixel sizes of $\mathcal{O}$(10 $\mu$m), therefore will be a powerful system for development of devices targeting precision position resolutions.


\vskip2pc

\section*{Acknowledgments}

We would like to thank personnel of the ELPH accelerator, Tohoku University, and the Fermilab Test Beam Facility  for providing beams in excellent condition throughout the experiment.
Expert advice and suggestions provided by Prof. Yasuo Arai and Dr. Toshinobu Miysohi, designers of INTPIX4, and by Dr. Ryutaro Nishimura, DAQ system developer, were valuable for conducting this study. We also thank Mr. Kyoya Misumi of Miyazaki University for helping in data acquisition  at the ELPH.
The current study was supported by the Tomonaga Center for the History of the Universe (TCHoU) of Universiy of Tsukuba and KAKENHI (19H00692).

\vspace{0.2cm}
\noindent


\let\doi\relax


\begin{thebibliography}{9}

\bibitem{1}
ELPH homepage,
\url{https://www.lns.tohoku.ac.jp/en/}.

\bibitem{2}
K. Hanagaki, High Energy News, {\bf 39-2} (2020) (in Japanese),
\url{https://www.jahep.org/hepnews/2020/Vol39_No2_4_TBL.pdf}.


\bibitem{3}
Y.~Arai, ``INTPIX4 User's Manual'', 
\url{https://soipix.jp/content/files/documents/INTPIX4doc\_v032.pdf}

\bibitem{4}
R.~Nishimura et al., ``DAQ Development for Silicon-On-Insulator Pixel Detectors", Proc. of  Int. 
Workshop on SOI Pixel Detector (SOIPIX2015), June 3-6, 2015, Sendai, Japan;,
\url{https://arxiv.org/abs/1507.04946}.



\bibitem{5}
H.~Yamauchi,``Performance of a tracking system constructed using large area INTPIX4 SOI pixel detectors evaluaed in 120 GeV beam (in Japanese)'',  Master thesis, University of Tsukuba (2019).

\bibitem{6}

H. Hayashi et al.,  Nucl. Instr. Methods A924, 400--403 (2019),
\url{https://doi.org/10.1016/j.nima.2018.09.042}.

\bibitem{7}
T.~Ishikawa, ``Electron $\cdot$ positron beamline V (in Japanese)'', HD no. 440$\mathcal{E}$, Nov 2020.

\bibitem{8}
K.~Hara et al., ``Development of Silicon-on-Insulatpr Devices'', PoS(Vertex2017) 035,  2018,
\url{https://doi.org/10.22323/1.309.0035}.


\bibitem{9}
H. Murayama et al., Nucl. Instr. Methods A978, 164417 (2020),
\url{https://doi.org/10.1016/j.nima.2020.164417}.

\bibitem{10} ILC homepage,
\url{https://linearcollider.org}.


\bibitem{11}
J.~Allison et al., ``Geant4 developments and applications'', IEEE TNS 53-1, 270--278 (2006),
\url{https://ieeexplore.ieee.org/document/1610988}.


\bibitem{12}
H.~Jansen et al., arXiv:1603.09669v2 [physics.ins-det] (2016),
.\url{https://doi.org/10.48550/arXiv.1603.09669}

\end{thebibliography}

\end{document}